\documentclass{kluwer}    

\usepackage[]{graphicx}
\usepackage[]{psfig}
\usepackage[]{epsf}

\newdisplay{guess}{Conjecture}

\begin{document}                                                              
\begin{article}
\begin{opening}         
\title{Abundances from Disentangled Component Spectra of Close Binary 
    Stars: An Observational Test of an Early Mixing in High-Mass Stars} 

\author{K. \surname{Pavlovski}}
\institute{Department of Physics, University of Zagreb, Zagreb, Croatia}
\author{D. E. \surname{Holmgren}}
\institute{SMART Technologies, Inc., Calgary, Canada}
\author{P. \surname{Koubsk\'{y}}}
\institute{Astronomcal Institute of the Academy of Sciences, Ond\v{r}ejov, Czech Republic}
\author{J. \surname{Southworth}}
\institute{Niels Bohr Institute, Copenhagen University, Denmark}
\author{S. \surname{Yang}}
\institute{Department of Physics \& Astronomy, University of Victoria, Canada}

\runningauthor{Pavlovski et al.}
\runningtitle{Abundances from disentangled component spectra}
\date{September 30, 2005}

\begin{abstract}
Recent theoretical calculations of stellar evolutionary tracks for
rotating high-mass stars suggests that the chemical composition of
the surface layers changes even whilst the star is evolving on the
Main Sequence.  The abundance analysis of binary components with
precisely known fundamental stellar quantities allows a powerful
comparison with theory. The observed spectra of close binary stars 
can be separated into the individual spectra of the component stars 
using the method of spectral disentangling on a time-series of 
spectra taken over the orbital cycle. Recently, Pavlovski \& 
Hensberge (2005, A\&A, 439, 309) have shown that, 
even with moderately high line-broadening, metal abundances can be 
derived from disentangled spectra with a precision of 0.1 dex. In a 
continuation of this project we have undertaken a detailed abundance 
analysis of the components of another two high-mass binaries, V453 
Cyg, and V380 Cyg. Both binaries are well-studied systems with 
modern solutions. The components are close to the TAMS and therefore very 
suitable for an observational test of early mixing in high-mass 
stars.
\end{abstract}
\keywords{stars: abundances  -- stars: binaries: eclipsing -- stars: 
binaries: close -- stars: binaries: spectroscopic }

\end{opening}           

\section{Introduction}  

New stellar evolution models which include the effects of rotationally 
induced mixing (Heger \& Langer 2000; Meynet \& Maeder 2000) have
considerably changed our understanding of the evolution of high-mass
stars, particularly during the early phases of core hydrogen burning.
Rotation is now recognized as an important physical effect which 
substantially changes the lifetimes, chemical yields and stellar 
evolution. Theoretical predictions can be observationally tested, 
and some attempts at this have already been made (c.f.~Venn et al.\ 2002). 

Chemical analysis of the components of binary stars with precisely 
known fundamental stellar parameters allows a powerful comparison 
with theory. However, the precision of empirical abundances from 
doub\-le-lined binaries is hampered by increased line blending and by 
dilution of the spectral lines in the composite spectra. The 
techniques of spectral disentangling (Simon \& Sturm 1994; Hadrava 
1995) and Doppler tomography (Bagnuolo \& Gies 1991) overcome these 
difficulties by separating the spectra of the individual components 
contained in a time-series of composite spectra taken over the 
orbital cycle.

\begin{figure}
\begin{tabular}{ll}
\includegraphics[width=5.7cm]{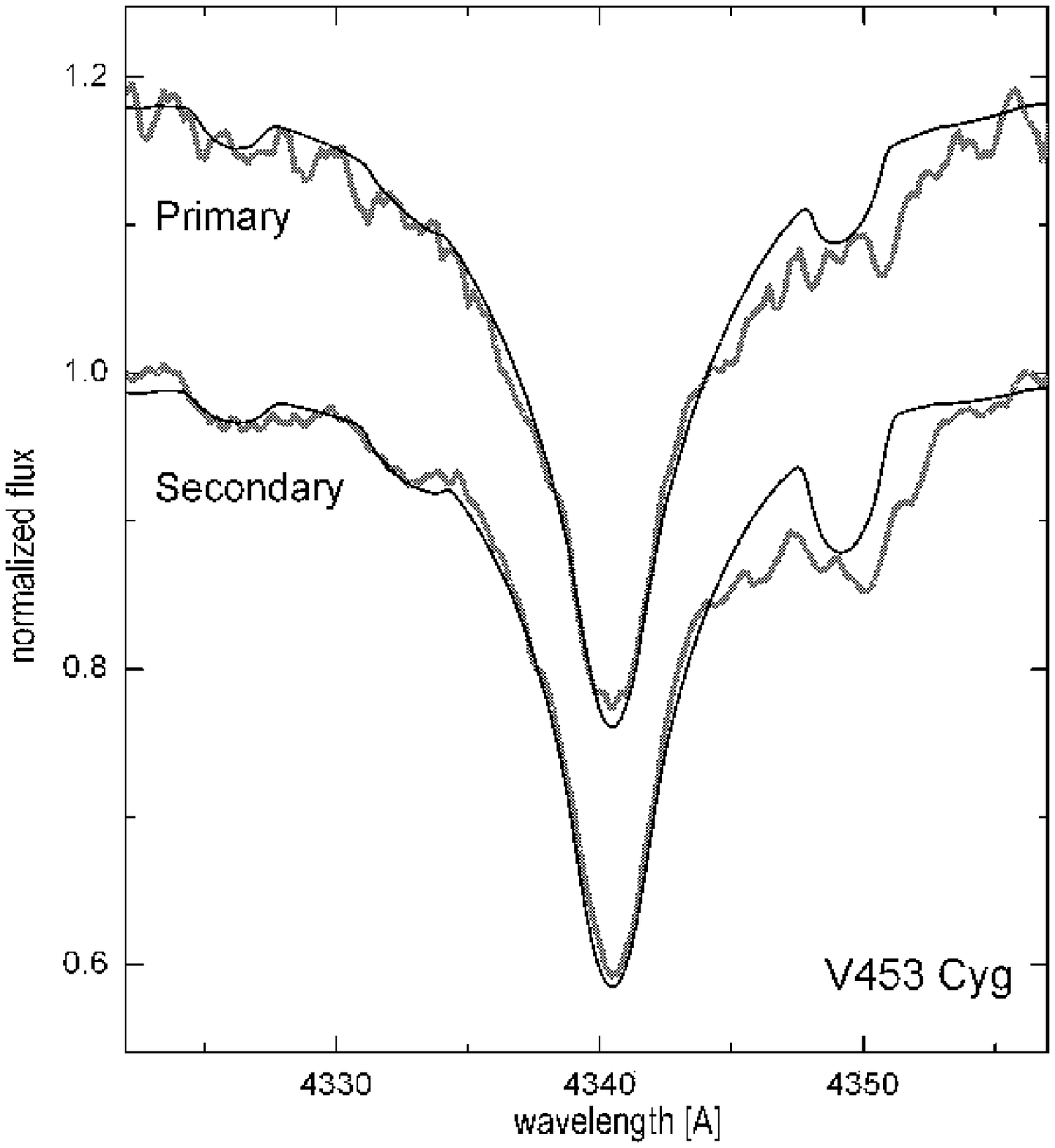} &
\includegraphics[width=5.7cm]{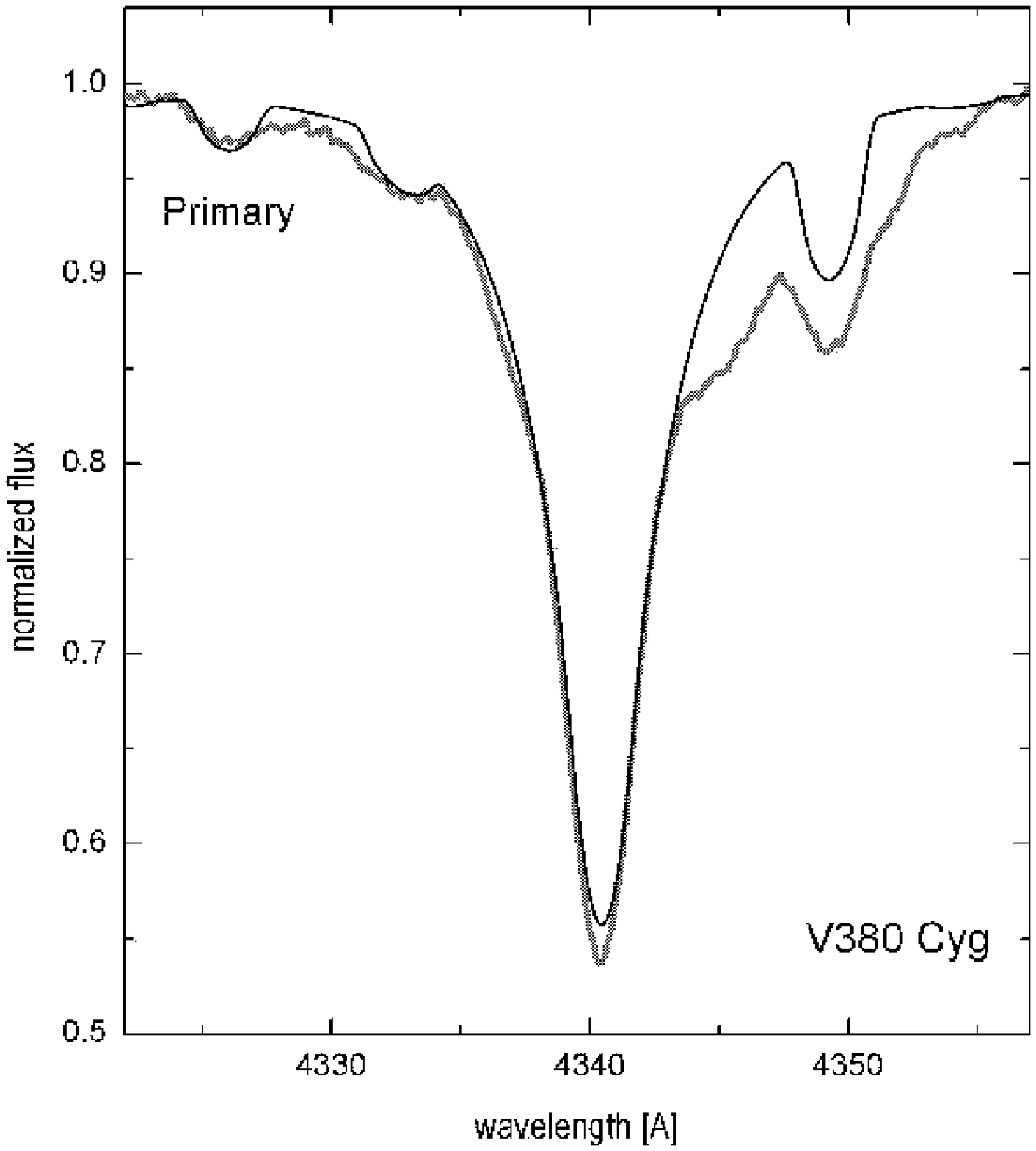}
\end{tabular}

\caption{The best fit of the calculated profiles (thin black line) of the 
H$\gamma$ line compared to the observed profiles (thick gray) for the
components of V453 Cyg, left panel, and the primary component
of V380 Cyg, right panel.}
\end{figure}

Pavlovski \& Hensberge (2005) have performed a detailed spectral line 
analysis of disentangled component spectra of the eclipsing early-B 
binary V578 Mon in the open cluster NGC~2244, which is embedded in the 
Rosette Nebula. It is based on the disentangled spectra obtained by 
Hensberge, Pavlovski \& Verschueren (2000) when deriving the orbit and 
the fundamental stellar parameters of this eclipsing, detached, 
double-lined system. V578 Mon consists of very young ($2.3\pm0.2 
\times 10^6$ yr) high-mass stars, $M_A = 14.54\pm 0.08$ M$_\odot$ and 
$M_B = 10.29 \pm 0.06$ M$_\odot$. The stars rotate moderately fast 
($v \sin i \sim 100$ km$\,$s$^{-1}$). By comparison with spectra of
single stars in the same open cluster (Vrancken et al.\ 1997),
temperature-dependent, faint spectral features are shown to reproduce
well in the disentangled spectra, which validates a detailed 
quantitative analysis of these component spectra. An abundance
analysis differential to a sharp-lined single star, as applied
earlier in this cluster to single stars rotating faster than the
components of V578 Mon, revealed abundances in agreement with the
cluster stars studied by Vrancken et al.\ (1997) and the large
inner-disk sample of Daflon et al.\ (2004). Pavlovski \& Hensberge 
(2005) have concluded that methods applicable to observed single
star spectra perform well on disentangled spectra, given that the
latter are carefully normalised to their intrinsic continua. 

Since the fundamental stellar and atmospheric parameters of eclipsing, 
double-lined spectroscopic binaries  are known with much better 
accuracy than in the case of single stars, the comparison with 
evolutionary models can be more direct and precise. The present work 
is a continuation of an observational project to test rotationally 
induced mixing in high-mass stars from disentangled component spectra 
of close binary stars.

We will now present preliminary results on two interesting early-B 
type systems, V453 Cyg and V380 Cyg. Both systems are detached, 
eclipsing, double-lined spectroscopic binaries and have reliable 
modern absolute dimensions, published by Southworth et al.\ (2004)
for V453 Cyg and Guinan et al.\ (2000) for V380 Cyg (Table I).

\begin{table}[!t] 
\begin{tabular}{lrrrr} \hline
Qnty/Comp               & V453 Cyg A$^1$  & V453 Cyg B$^1$  & V380 Cyg A$^2$  & V380 Cyg B$^2$  \\ 
\hline
$M$ [M$_{\odot}]$       & $14.36\pm0.20$  & $11.11\pm0.13$  & $11.1\pm0.5$    & $6.95\pm0.25$   \\
log $g$ [cgs]           & $3.731\pm0.012$ & $4.005\pm0.015$ & $3.148\pm0.023$ & $4.133\pm0.023$ \\
$T_{\rm eff}$ [K]       & $26\,600\pm500$ & $25\,500\pm800$ & $21\,350\pm400$ & $20\,500\pm500$ \\
$v \sin i$              & $107\pm9$       & $97\pm20$       & $98\pm4$        & $32\pm6$        \\
$\epsilon_{\rm He}$$^3$ & $0.13\pm0.01$   & $0.09\pm0.01$   & $0.14\pm0.01$   &  --             \\ 
\hline \end{tabular}
\caption[]{Fundamental parameters for the stars in V453 Cyg and V380 Cyg.}
Notes: (1) Southworth et al.\ (2004); (2) Guinan et al.\ (2000); (3) This work.
\end{table}

\begin{figure}
\begin{tabular}{cc}
\includegraphics[width=5.7cm]{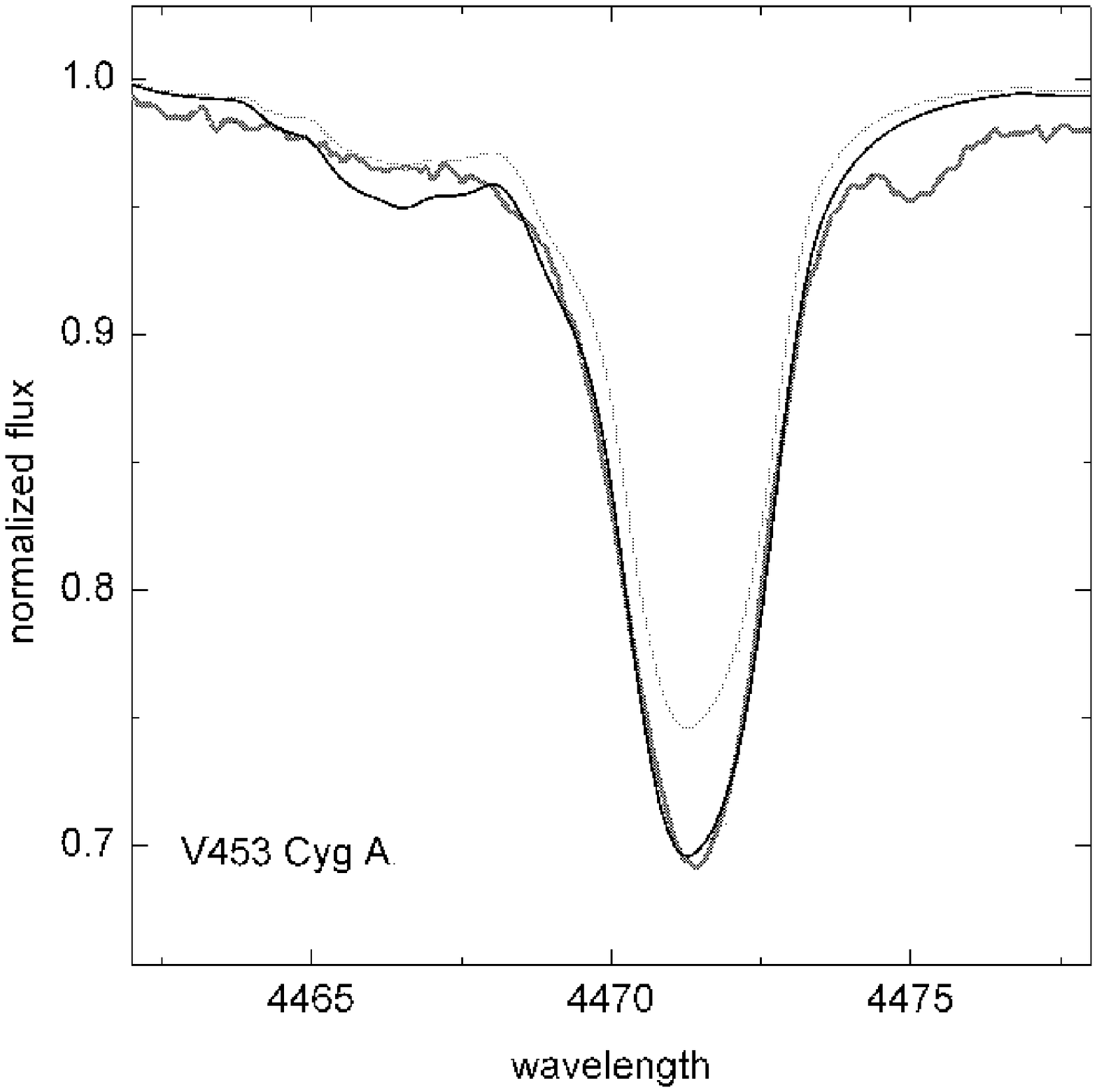} &
\includegraphics[width=5.7cm]{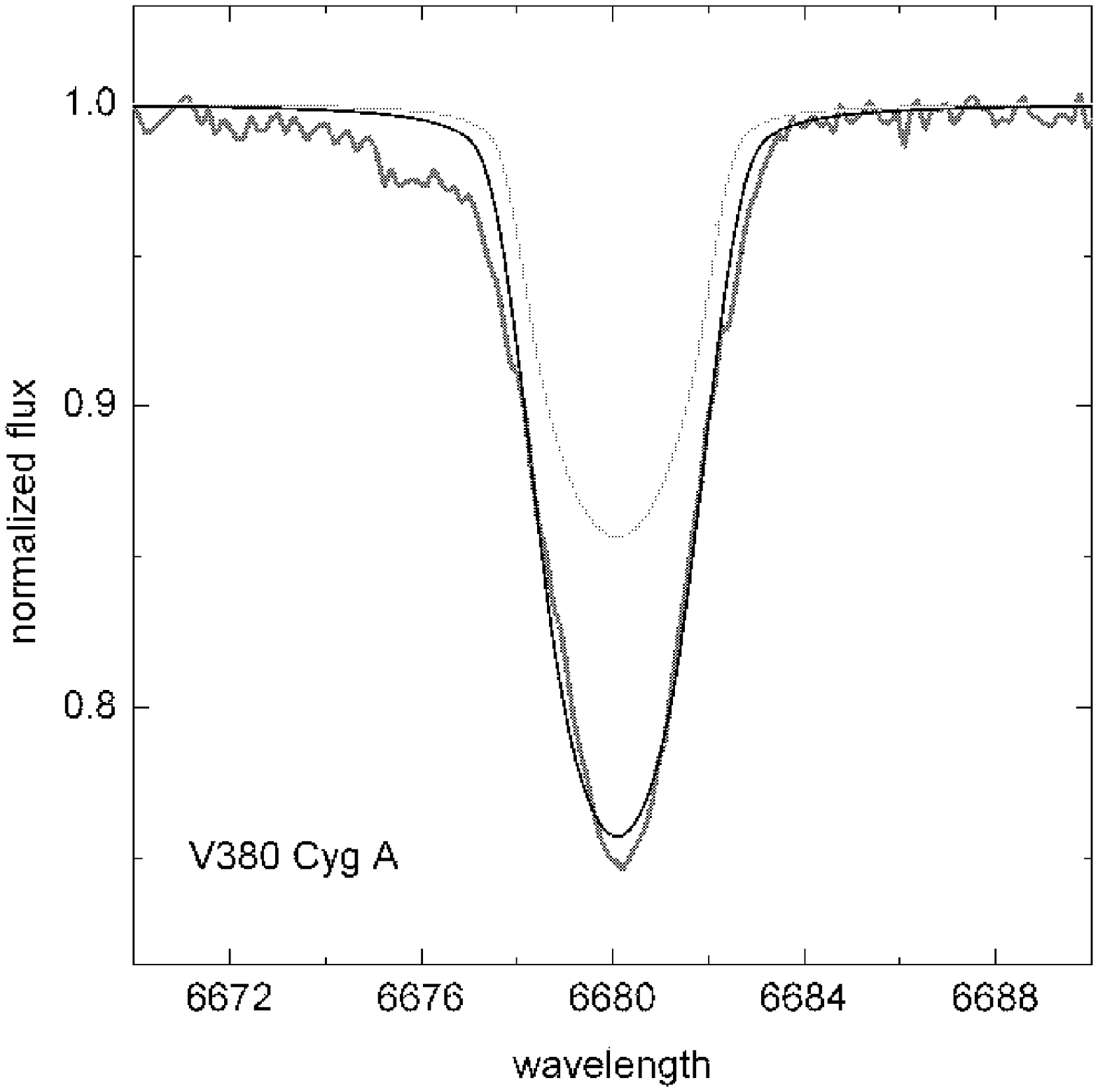}
\end{tabular}
\caption{The best fitting calculated profiles (thin black) of the He~{\sc i} 
4471 {\AA}  and 6678 {\AA} lines compared to the observed profiles 
(gray) in the primary component of V453 Cyg (left panel), and the 
primary component of V380 Cyg (right panel). Light gray lines represent
profiles for the solar helium abundance.} 
\end{figure}

\section{Spectroscopy and Method}

Several different sets of spectra were obtained for both binaries. We 
will briefly describe these observations.

{\em V453 Cyg}: This binary was observed in 1991 and 1992 with the 2.2-m
telescope at German-Spanish Astronomical Center on Calar Alto, Spain. 
Four spectral windows were observed with the coud\'e spectrograph. A 
total of 28 spectra were collected. These spectra were kindly put at 
our disposal by Dr. Klaus Simon. Further description can be found in 
Simon \& Sturm (1994). Another similar set, in two spectral windows, 
was secured by one of the authors (JS) in 2001 with the 2.5-m Isaac Newton
Telescope at La Palma (Southworth et al.\ 2004). A total of 41 spectra 
were obtained. An additional set of six spectra in the red region centred 
on H$\alpha$ were secured by DH on the 1.2-m telescope at the DAO in 2001.

{\em V380 Cyg}: Eight spectra centred on H$\gamma$ were obtained by PK 
and KP at the coud\'e spectrograph on the 2-m telescope in Ond\v{r}ejov 
in 2004. An additional two spectra in the same region were obtained by 
PK on the 1.2-m telescope at the DAO, Victoria, also in 2004. An 
additional set of eight red spectra centred on H$\alpha$, from the same 
telescope, were obtained by SY in 2002 and are also used here.


To isolate the individual spectra of both components in V453 Cyg we 
have made use of the spectral disentangling technique (Simon \& Sturm 
1994, Hadrava 1995). The computer codes {\sc FDBinary} (Iliji\'{c} et 
al.\ 2004) and {\sc cres} (Iliji\'{c} 2004), which rely on the Fourier 
transform technique (Had\-ra\-va 1995), and the SVD technique in 
wavelength space (Simon \& Sturm 1994), respectively, were used.
Spectral disentangling is a powerful method which has found a variety
of applications in close binary research (c.f.~Holmgren et al.\ 1998; 
Hensberge et al.\ 2000; Harries et al.\ 2003; Harmanec et al.\ 2004).

The non-LTE line-formation calculations are performed using 
{\sc Detail} and {\sc Surface} (Butler \& Giddings 1985). However, 
hydrostatic, plane-parallel, line-blanketed LTE model atmospheres 
calculated with the {\sc Atlas9} code (Kurucz 1983) have also been 
used. This hybrid approach has been compared with the state-of-the-art 
non-LTE model atmosphere calculations and excellent agreement has been 
found for the hydrogen and helium lines (Przbylla 2005).

\begin{figure}
\centerline{\includegraphics[width=7.0cm]{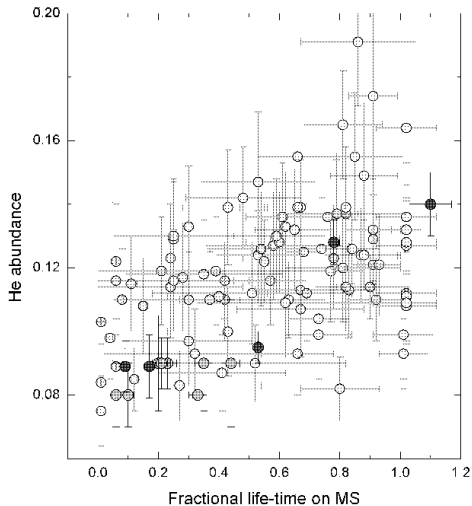}}

\caption{Abundances of helium in the components of the close binaries 
(filled symbols; dark symbols show the results of this work) 
overplotted on the results for large sample of single early-B 
type stars (open symbols) of Lyubimkov et al.\ (2004).}
\end{figure}

\section{Results and Conclusion}

In the observed spectral ranges the helium abundance can be derived only
from the lines centred at 4378, 4471, 4718 and 6678 \AA. As discussed
by Lyubimkov et al.\ (2004), calculations for He I 4378 {\AA} are less
reliable using {\sc detail} since only transitions up to level $n = 4$
are considered explicitly. Since level populations can be affected by
the microturbulent parameter $V_{turb}$ it should be also included in
calculations and adjusted to the observed line profiles. 

First, a check and slight adjustment of the effective temperature to 
the individual component spectra for V453 Cyg, and the primary of V380 
Cyg, has been made. As an example, fitting of the calculated to the 
observed line profiles of the H$\gamma$ line is shown in Fig.\ 1. A 
simultaneous fit of the helium abundance $\epsilon_{\rm He}$, and 
microturbulent velocity $V_{turb}$ has then been performed from the 
grid of the calculated spectra, while $T_{\rm eff}$ and $\log g$ have 
been kept fixed. 

The helium enrichment has been found for the primary component in the
system V380 Cyg by Lyubimkov et al.\ (1996). The helium abundance they
derived, $\epsilon_{\rm He} = 0.19\pm0.05$, is considerably larger than
the value derived in the present work. The complete analysis and
discussion of possible sources of the discrepancy will be published
elsewhere. 

Recently, Lyubimkov et al.\ (2004) have derived the helium abundances
in a large sample of early-B type stars. Their results are plotted in 
Fig.\ 3 as open symbols. This confirmed their finding that helium is 
becoming enriched in high-mass stars already on the main sequence. 
However, due to large errors in deriving the fundamental parameters 
for the single stars, there is considerable scatter in their diagram. 
Overplotted by filled symbols are results for the components of 
eclipsing, double-lined spectroscopic binaries, in light-gray 
(c.f.~Pavlovski 2004). In dark-gray are presented the results of this 
work. The general finding that in the later phases on the main 
sequence helium is enriched is confirmed, but there is disagreement 
for early phases for which results from the close binaries are very 
consistent and are giving a helium abundance close to the solar 
value. However, the sample is still rather limited and more work 
is needed to have complete picture on the helium enrichment on the MS 
for high-mass stars.

\end{article}
\end{document}